\documentclass[8.5pt,twoside,twocolumn]{article}
\usepackage[super,sort&compress,comma]{natbib} 
\usepackage{mhchem}
\usepackage{times,mathptmx}
\usepackage{sectsty}
\usepackage{color} 
\usepackage{balance} 

\usepackage{graphicx} 
\usepackage{lastpage}
\usepackage[format=plain,justification=raggedright,singlelinecheck=false,font=small,labelfont=bf,labelsep=space]{caption} 
\usepackage{fancyhdr}
\newcommand{\clr}[1]{\textcolor{black}{#1}}
\begin{document}

\thispagestyle{plain}
\fancypagestyle{plain}{
\renewcommand{\headrulewidth}{1pt}}
\renewcommand{\thefootnote}{\fnsymbol{footnote}}
\renewcommand\footnoterule{\vspace*{1pt}%
\hrule width 3.4in height 0.4pt \vspace*{5pt}} 
\setcounter{secnumdepth}{5}

\makeatletter 
\def\subsubsection{\@startsection{subsubsection}{3}{10pt}{-1.25ex plus -1ex minus -.1ex}{0ex plus 0ex}{\normalsize\bf}} 
\def\paragraph{\@startsection{paragraph}{4}{10pt}{-1.25ex plus -1ex minus -.1ex}{0ex plus 0ex}{\normalsize\textit}} 
\renewcommand\@biblabel[1]{#1}            
\renewcommand\@makefntext[1]%
{\noindent\makebox[0pt][r]{\@thefnmark\,}#1}
\makeatother 
\renewcommand{\figurename}{\small{Fig.}~}
\sectionfont{\large}
\subsectionfont{\normalsize} 

\fancyfoot{}
\fancyfoot[RO]{\footnotesize{\sffamily{1--\pageref{LastPage} ~\textbar  \hspace{2pt}\thepage}}}
\fancyfoot[LE]{\footnotesize{\sffamily{\thepage~\textbar\hspace{3.45cm} 1--\pageref{LastPage}}}}
\fancyhead{}
\renewcommand{\headrulewidth}{1pt} 
\renewcommand{\footrulewidth}{1pt}
\setlength{\arrayrulewidth}{1pt}
\setlength{\columnsep}{6.5mm}
\setlength\bibsep{1pt}

\twocolumn[
  \begin{@twocolumnfalse}

\noindent\LARGE{\textbf{Formation of the 
prebiotic molecule NH$_2$CHO on astronomical amorphous solid water surfaces:  
accurate tunneling rate calculations}}

\vspace{0.6cm}

\noindent\large{\textbf{Lei Song,\textit{$^{a}$} Johannes K\"astner,\textit{$^{a\ddag}$}}}\vspace{0.5cm}

\noindent\textit{\small{\textbf{Received Xth XXXXXXXXXX 20XX, Accepted Xth XXXXXXXXX 20XX\newline
First published on the web Xth XXXXXXXXXX 200X}}}

\noindent \textbf{\small{DOI: 10.1039/b000000x}}
\vspace{0.6cm}

\noindent\normalsize{Investigating how formamide forms in the interstellar
  medium is a hot topic in astrochemistry, which can contribute to our
  understanding of the origin of life on Earth.  We have constructed a QM/MM
  model to simulate the hydrogenation of isocyanic acid on amorphous solid
  water surfaces to form formamide. The binding energy of HNCO on the ASW
  surface varies significantly between different binding sites, we found
  values between $\sim$0 and 100 kJ~mol$^{-1}$. The barrier for the
  hydrogenation reaction is almost independent of the binding energy,
  though. We calculated tunneling rate constants of H + HNCO $\rightarrow$
  NH$_2$CO at temperatures down to 103~K combining QM/MM with instanton
  theory.  Tunneling dominates the reaction at such low temperatures.  The
  tunneling reaction is hardly accelerated by the amorphous solid water
  surface compared to the gas phase for this system, even though the
  activation energy of the surface reaction is lower than the one of the gas-phase
  reaction.  Both the height and width of the barrier affect the tunneling
  rate in practice. Strong kinetic isotope effects were observed by comparing
  to rate constants of D + HNCO $\rightarrow$ NHDCO. At 103~K we found a KIE
  of 231 on the surface and 146 in the gas phase. Furthermore, we investigated
  the gas-phase reaction NH$_2$ + H$_2$CO $\rightarrow$ NH$_2$CHO + H and
  found it unlikely to occur at cryogenic temperatures. The data of our
  tunneling rate constants are expected to significantly influence
  astrochemical models.}
\vspace{0.5cm}
 \end{@twocolumnfalse}
  ]

\section{Introduction}
\footnotetext{\dag~Electronic Supplementary Information (ESI) available:
  geometric details, lists of calculated rate constants. See DOI:
  10.1039/b000000x/}

\footnotetext{\textit{Institute for Theoretical Chemistry, University of
    Stuttgart, Pfaffenwaldring 55, 70569 Stuttgart, Germany, kaestner@theochem.uni-stuttgart.de}}

Formamide (NH$_2$CHO), the simplest molecule containing a peptide bond, has
attracted much attention in the field of astrochemistry owing to its potential
role as a prebiotic precursor in the origin of life on Earth.  It was first
detected in a molecular cloud in 1971 by Rubin et al.~\cite{rubin:71}.  Since
then, formamide has been found on comets and in a variety of star-forming
regions, such as in high mass young stellar objects (YSOs),\cite{bisschop:07}
outflow shock regions,\cite{yama:12, mendoza:14} and on the comet
Hale--Bopp.\cite{bockelee:00} Recently, L\'{o}pez-Sepulcre et
al. \cite{lopez:15b} detected NH$_2$CHO in five out of ten low- and
intermediate-mass pre-stellar and protostellar objects as well as isocyanic
acid (HNCO) in all ten sources under study. They found a tight and almost
linear correlation between NH$_2$CHO and HNCO abundance, which indicates the
existence of a chemical relation between those two molecules.

The formation sequence for complex organic molecules like NH$_2$CHO can occur
either in gas-phase or on the surface of dust grains in the interstellar
medium.\cite{raunier:04, garrod:08, redondo:14} Consecutive hydrogenations of
HNCO on the mantles of dust grains were proposed as a likely formation route to
produce NH$_2$CHO:
\begin{equation}
 \ce{H + HNCO -> NH2CO}
 \label{re:nh}
\end{equation}
\begin{equation}
 \ce{H + NH2CO -> NH2CHO}
 \label{re:nh2}
\end{equation}
Since (\ref{re:nh2}) is a radical-radical recombination reaction it is
barrierless. Reaction (\ref{re:nh}) is rate-limiting and thus the focus of 
this study will be on it. Nguyen et al.~\cite{nguyen:96} investigated
(\ref{re:nh}) in the gas phase and suggested the NH$_2$CO radical as
the primary intermediate and NH$_2$+CO as the fragment
products. However, a surface can dissipate the extra energy on the
NH$_2$CO radical and, thus, stabilize it.  However, in recent experimental
work by Noble et al.,\cite{noble:15} the low temperature reaction of
solid phase HNCO with H atoms did not produce detectable amounts of
NH$_2$CHO. Even though, formation of NH$_2$CHO from HNCO could be
possible on other surfaces, like amorphous solid water (ASW) surfaces.

A gas-phase formation route of NH$_2$CHO was investigated by Barone et
al.~\cite{barone:15} using quantum chemical computations. They suggested the
reaction
\begin{equation}
  \ce{NH2 + H2CO -> NH2CHO + H}
  \label{re:h2co}
\end{equation}
to be barrierless and therefore a viable route for NH$_2$CHO-formation in the
gas phase. We will briefly address this reaction in the present work as
well. 

The increased concentration of active species on the surface of dust
grains lends weight to the surface formation route. The mantles of dust grains
are predominantly composed of H$_2$O in the amorphous phase combined with
other molecules such as CO, CH$_4$, NH$_3$, and traces of other molecules like
HNCO, and NH$_2$CHO. Therefore, modeling the reactions on an ASW surface is
probably close to the astronomical environment.\cite{hama:13} The
temperature is always low on the ASW surface, where quantum tunneling is
expected to play an important role in chemical reactions.  In addition,
quantum tunneling is also very likely to happen in the hydrogenation reactions
owing to the light reactant H atoms.\cite{mei16}

In this work we study reaction (\ref{re:nh}) on an ASW surface using
hybrid quantum mechanics/molecular mechanics (QM/MM)
calculations. Combined with instanton theory, we provide tunneling
rates of this reaction in the gas phase and on the ASW surface.

\section{Methods}
\label{sec:meth}

\subsection{System preparation}

The ASW surface was prepared by classical molecular dynamics (MD) simulations
with NAMD.\cite{phillips:05} The initial sample is produced by VMD version
1.9.2~\cite{hum96a} containing 9352 TIP3P water\cite{jor83} molecules. These
were simulated in a slab of 85 {\AA} $\times$ 85 {\AA} and a thickness of
approximately 36 {\AA}.  Periodic boundary conditions were applied along all
three Cartesian axes with about 70 {\AA} of vacuum between the slabs.  This
system was treated as a canonical ensemble, equilibrated at 300~K using a
Langevin thermostat for 100~ps. After that, the thermostat was instantaneously
quenched to 10~K and the system was left for 20~ps to produce a thermally
equilibrated bulk amorphous water at low temperature. A hemisphere with a
radius of 34 {\AA} was cut out of the slab to be used in the following QM/MM
calculations.

A large sample of different binding sites on the surface was generated. The
HNCO molecule was placed at 113 positions on a regular 2D-grid with a step
size of 2~{\AA} covering a circular area with a radius of 12~{\AA}. In each of
the 113 points, the molecule was placed 2~{\AA} above the surface. Water
molecules with at least one atom within 6~{\AA} were treated by QM (typically
about 23 molecules), water molecules within 12~{\AA} were optimized (typically
about 161). All other molecules of the hemispheric model were frozen.

\subsection{QM/MM method}
Both geometry optimization and tunneling rate calculations were
performed using a state-of-art QM/MM approach.\cite{warshel:72,
  warshel:76} In this approach, the reactants H, HNCO and their closer
water surroundings were treated with density functional theory (DFT)
while more distant water molecules were described by the TIP3P force
field.

The hybrid QM/MM calculations\cite{warshel:72, warshel:76} were carried out
with ChemShell,\cite{she03,met14} using an additive electrostatic embedding
scheme, where the MM point charges polarize the QM electron density. 
We used B3LYP~\cite{ste94}/def2-SVPD~\cite{rappoport:10} to calculate the binding
energies and binding site geometries. 
Different density functionals were tested and compared to coupled cluster reference
values as outlined in Section~\ref{sec:ben}. 
On the basis of this comparison, 
BHLYP-D3~\cite{becke:93b,lee:88,grimme:10}/def2-TZVP~\cite{florian:05} was used for
barriers and rate calculations.  The quantum chemical program package
TURBOMOLE 6.6\cite{turbomole} was used for the QM part while
DL\_POLY\cite{smith:02} built into ChemShell, was used for MM part. Force
field parameters for H and HNCO (only the van-der-Waals parameters are used in
QM/MM) were chosen in analogy to the CHARMM22 force
field.\cite{mackerell:98,mackerell:00,mackerell:04} The open-source optimizer
DL\_FIND\cite{kaestner:09} was employed for geometry optimizations including
the search for binding sites, the search for transition states with the dimer
method\cite{hen99,hey05,kae08} and the determination of instanton paths using a modified
Newton--Raphson approach.\cite{rom11,rom11b}


\subsection{Instanton theory}

Tunneling rates in this work were calculated using instanton
theory\cite{lan67,mil75,col77,cal77,alt11,ric16} in its semiclassical
formulation.\cite{aff81,col88,han90, ben94,mes95,ric09,alt11,rom11,rom11b,zha14}
Instanton theory is based on statistical thermodynamics for the rate
expression in which the partition function from a quantum mechanical ensemble
is expressed via a Feynman path integral. Generally, this theory is only
applicable below the crossover temperature $T_\text{c}$:\cite{gillan:87}
\begin{equation}
 T_\text{c} = \frac{\hbar \omega_\text{b}}{2\pi k_\text{B}}
\end{equation}
where $\omega_\text{b}$ stands for the absolute value of the classical
imaginary frequency at the transition state, $k_\text{B}$ for the Boltzmann
constant and $\hbar$ for the reduced Planck constant.  At a given temperature
below $T_\text{c}$, the instanton itself is the tunneling path with the
highest statistical weight, which can be located using standard approaches for
finding transition states.\cite{rom11,rom11b}
Integrating along this path and combining it with the partition function of
reactant state, we can calculate instanton rate constants which consider
quantum tunneling effects. Due to its semi-classical nature, instanton theory
can offer a reasonable ratio of accuracy versus computational cost,
appropriate for our reactions with organic molecules on the ASW surface.
Instanton theory is meanwhile frequently used to calculate reaction rates in
different areas of chemistry.\cite{cha75,mil94,mil95,mil97,sie99,
  sme03,qia07,and09, gou10a,gou11,gou11b,
  rom11,gou10,jon10,mei11,gou11a,ein11,rom12,
  kry12,kae13,alv14,kae14,kry14,mei16}

The Feynman paths were discretized to 40 images at $T\ge 135$ K and 78 images
at lower temperature. Convergence was checked rigorously, e.g. at 100~K
doubling the number of images changed the rate constant by only 2\%.

\begin{figure}[h]
\centering
  \includegraphics[height=6.8cm]{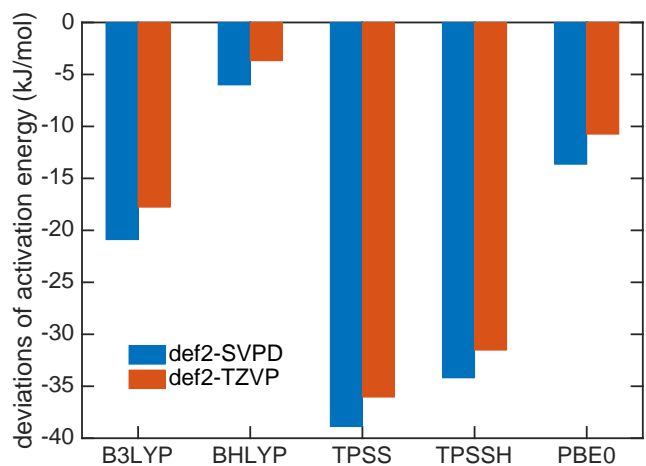}
  \caption{Deviations of activation energies of reaction (\ref{re:nh}) at
    different DFT levels with D3 dispersion correction from the results at
    UCCSD(T)-F12/cc-pVTZ-F12 level.}
  \label{fgr:benchmark}
\end{figure}

In order to make our calculated rate constants accessible to astromodellers, we
fitted them to a rate expression proposed previously:\cite{zhe10}
\begin{equation}
  k(T)=\alpha \left(\frac{T}{300\text{K}}\right)^\beta \exp\left(-\frac{\gamma(T+T_0)}{T^2+T_0^2}\right)
  \label{eq:zt}
\end{equation}
in which $\alpha$, $\gamma$ and $T_0$ were used as fitting parameters
and $\beta$ was set to one. 
\clr{The pre-exponential factor $\alpha$ has the same unit as the rate constant and
  can be interpreted as an attempt frequency. The parameter $\gamma$
  is related to the barrier height and $T_0$ is a temperature, which
  relates to the onset of strong tunneling. Any physical meaning of
  these fitting parameters should not be over interpreted, though.} 
Instanton rate constants were used for the
fit below $T_\text{c}$, rate constants calculated by transition state
theory with vibrations treated by quantum harmonic oscillators and a
symmetric Eckart barrier for tunnel corrections were used to fit above
$T_\text{c}$.  Eq.~(\ref{eq:zt}) describes classical thermal reactions
as well as tunneling rates with a single expression. For
$T_0\rightarrow 0$ it turns into the standard Arrhenius equation which
is used in many astrochemical models.

\section{Results}

\subsection{Benchmark calculations\label{sec:ben}}

Benchmark calculations were performed to choose a proper DFT functional for
the transition state search and tunnel rate calculations.  We calculated the
activation energy $E_\text{a}$ for reaction~(\ref{re:nh}) in the gas phase
based on B3LYP-D3~\cite{ste94,grimme:10}/def2-TZVPD~\cite{rappoport:10} optimized geometries using
UCCSD(T)-F12~\cite{adler:07,knizia:09}/cc-pVTZ-F12~\cite{peterson:08} 
on a RHF reference 
in MOLPRO 2012~\cite{MOLPRO}.  The
resulting $E_\text{a}$ of 32.7 kJ~mol$^{-1}$ was used as a reference and
compared to the data from B3LYP, BHLYP, TPSS, TPSSH and PBE0 functionals with
the def2-SVPD~\cite{rappoport:10} and def2-TZVP~\cite{florian:05} basis sets. All DFT calculations include D3 dispersion corrections~\cite{grimme:10}. The results are compared in
Fig.~\ref{fgr:benchmark}. The smallest deviation was found for the
BHLYP-D3~\cite{becke:93b,lee:88,grimme:10}/def2-TZVP~\cite{florian:05}
theory level which we selected as the proper quantum
mechanical level for QM molecules.

\begin{figure}
  \centering
  \includegraphics[width=8cm]{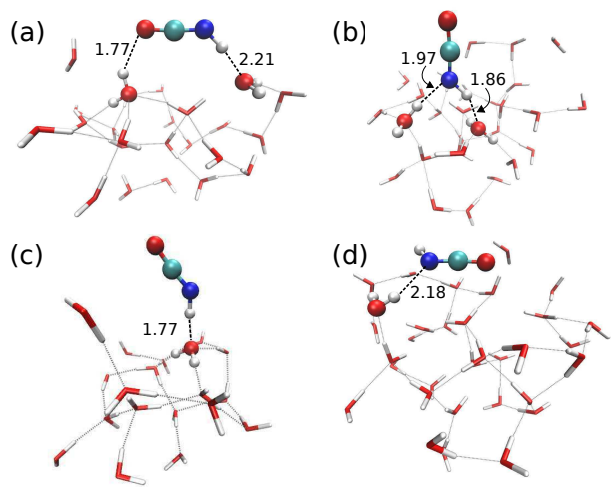}
  \caption{Four different HNCO binding modes on the amorphous solid water
    surface. Only QM molecules are shown, HNCO and all water molecules
    H-bonded to it are shown as ball-and-stick. Bond distances are given in
    {\AA}. }
  \label{fgr:binding_site}
\end{figure}

\subsection{HNCO binding sites and binding energies}

Reaction (\ref{re:nh}) originates from HNCO bound to the ASW surface. We
investigated different binding modes and their respective binding energies in
our QM/MM setup using the B3LYP~\cite{ste94}/def2-SVPD~\cite{rappoport:10} level for the QM calculations.
Geometry optimization was performed starting from 113 initial
structures. Among those, 90 jobs finished successfully and provided four types
of HNCO binding modes on the ASW surface as shown in Fig.~\ref{fgr:binding_site}.
Panel~(a) illustrates the major adsorption mode to which 48 out of the 90
cases belonged.  In this case the H and O ends of the HNCO molecule act as
H-bond donor and acceptor connecting to O and H atoms in the  water ice, respectively.
The N atom can also act as a H-bond acceptor while the H atom of the HNCO molecule
still serves as a H-bond donor to connect to an O atom from the water. This case
is depicted in Panel~(b) of Fig.~\ref{fgr:binding_site} and accounts for 34 of
90 cases. The remaining 8 cases resulted in binding modes 
where either the N atom or H atom in the HNCO molecule connects to
H or O of the surface, as shown in Panels~(c) and (d).

\begin{figure}[h]
\centering
  \includegraphics[height=6cm]{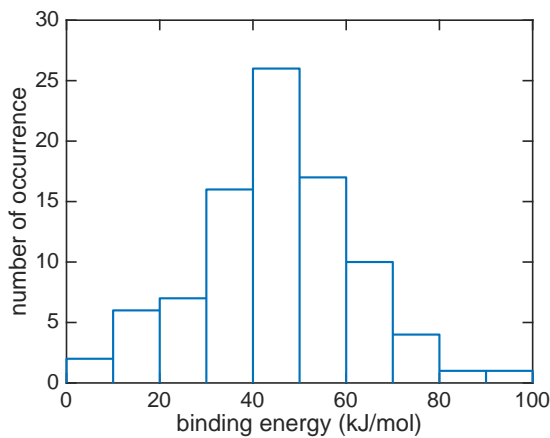}
  \caption{The distribution of HNCO binding energies on the amorphous solid
    water surface at the B3LYP~\cite{ste94}/def2-SVPD~\cite{rappoport:10} level of theory.}
  \label{fgr:binding_energy}
\end{figure}

The binding energy of HNCO on the ASW surface was 
the energy required to disassemble 
the \clr{adsorbed} HNCO from the surface into the gas phase.  
The minima of the ASW surface with and without HNCO in each of 
the 90 cases were calculated using the same QM, active and frozen water regions. 
Fig.~\ref{fgr:binding_energy} presents the distribution of binding energies
from the 90 cases. It is obvious that the binding energy is very broadly
distributed from 0 to about 100 kJ~mol$^{-1}$ with the largest fraction
between 40 and 50 kJ~mol$^{-1}$. The tighter bound sites are expected to be
occupied preferentially, which leads to a surface-coverage dependent binding
energy. No clear correlation can be found between the binding modes
distinguished in Fig.~\ref{fgr:binding_site} and the binding energies. The
rough surface of ASW leads to the significant spread of binding energies,
which likely is of relevance for astrochemical modeling of adsorption and
desorption processes. The binding energies are given in
Fig.~\ref{fgr:binding_energy} without considering the vibrational zero point
energy (ZPE). We calculated the ZPE for the four representative modes shown in
Fig.~\ref{fgr:binding_site}. They reduce the binding energy by 8.0, 5.4, 2.3,
and 7.7~kJ~mol$^{-1}$ for the modes a, b, c, and d, respectively. Thus, the
influence of the ZPE on binding is small.

\subsection{Transition states}

\begin{figure}[h]
\centering
  \includegraphics[height=7.2cm]{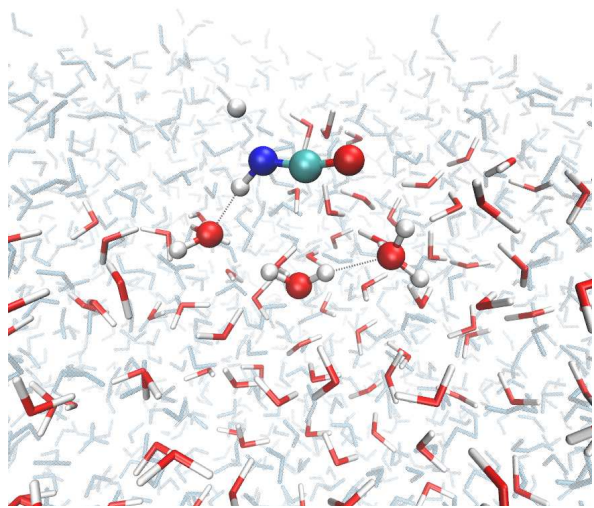}
  \caption{Optimized geometry of TS1 of reaction (\ref{re:nh}) on the ASW
    surface. In the TS search the QM region was restricted to the molecules
    shown as ball-and-stick models. All red/white water molecules were active,
    the blue/gray ones frozen.}
  \label{fgr:ts_003}
\end{figure}

We investigated transition states for four different binding geometries with
rather different binding energies. The resulting data are given in
Table~\ref{tbl:tss}. The transition structures are labeled TS1 to TS4. Their
binding energies differ between 27.9 and 80.3~kJ~mol$^{-1}$. The attack by a
hydrogen atom at the N-site of HNCO requires the latter to be
accessible. Thus, binding modes (a) and (c) of the ones depicted in
Fig.~\ref{fgr:binding_site} are most promising. TS1, TS3, and TS4 correspond
to binding mode (a) while TS2 corresponds to binding mode (c). For the
transition state search and the following tunneling rate calculations, we
restricted the QM region to H+HNCO plus just three water molecules (5 for TS2,
4 for TS4), see Fig.~\ref{fgr:ts_003}. While the same set of atoms (12~\AA)
was optimized as in the investigations of the binding sites, the Hessian
calculations were restricted to the QM region.

All data in Table~\ref{tbl:tss} refer to a reactant state with HNCO adsorbed
on the surface and H in the gas phase, i.e. to an Eley--Rideal-type (ER)
surface reaction mechanism.  Compared with the transition state in the gas
phase, the ones on the ASW surface have slightly lower activation energies
$E_\text{a}$. Without ZPE the four surface-bound activation energies are 3.9
to 0.7 kJ~mol$^{-1}$ lower than the gas-phase $E_\text{a}$, including ZPE they are
between 4.4 kJ~mol$^{-1}$ lower and 0.3 kJ~mol$^{-1}$ higher. Note that despite the large
spread in binding energies of the different adsorption sites, the associated
activation energies are very similar. This indicates similar rate constants,
which will be discussed in the following section. The N--H bond distances of
the transition states on the surface are generally slightly longer than in the gas
phase, see Table~\ref{tbl:tss}, indicating an earlier transition state on the
surface.

The transition states TS1, TS3, and TS4 describe a movement of the
hydrogen atom coming from the gas phase above the surface. By
contrast in TS2, which originates from a structure like the one in
Fig.~\ref{fgr:binding_site}~(c), the hydrogen atom approaches the
nitrogen site from closer to the surface, see also Fig. 1 of the
Supporting Material. In this case, a well-defined
pre-reactive minimum with H loosely bound to the surface was
found. This corresponds to a possible reactant site for a
Langmuir--Hinshelwood (LH) mechanism. The barrier with respect to the
LH reactant state is 34.6 kJ~mol$^{-1}$ (37.9 kJ~mol$^{-1}$ with ZPE).

\begin{table}[h]
\small
  \caption{Comparison of transition states in gas and on the amorphous solid
    water surface. The energies are given in kJ~mol$^{-1}$, frequencies in cm$^{-1}$,
    temperatures in K and bond distances in \AA.}
  \label{tbl:tss}
  \begin{tabular*}{\columnwidth}{@{\extracolsep{\fill}}lccccc}
    \hline
                        & gas & \multicolumn{4}{c}{ASW}    \\ \cline{3-6}
                        &  TS   &  TS1    &  TS2    &  TS3   & TS4    \\
    \hline
     HNCO binding energy &       & 48.1    & 27.9    & 80.3   & 52.1  \\
    N--H bond distance   & 1.542 & 1.546   & 1.532   & 1.546  & 1.547  \\
    $\omega_\text{b}$     & 1339i & 1240i   & 1271i   & 1268i  & 1262i  \\
    $E_\text{a}$ (ER mechanism) & 30.6  & 26.7    & 29.9    & 28.4   & 27.9   \\
    $E_\text{a}$ incl. ZPE& 36.2    & 31.8    & 36.5   & 32.7  & 32.7  \\
    $T_\text{c}$          & 307   &  284    &  291    & 290    & 289    \\
    \hline  
    
  \end{tabular*}
\end{table}

\begin{figure}[h]
\centering
  \includegraphics[height=6.5cm]{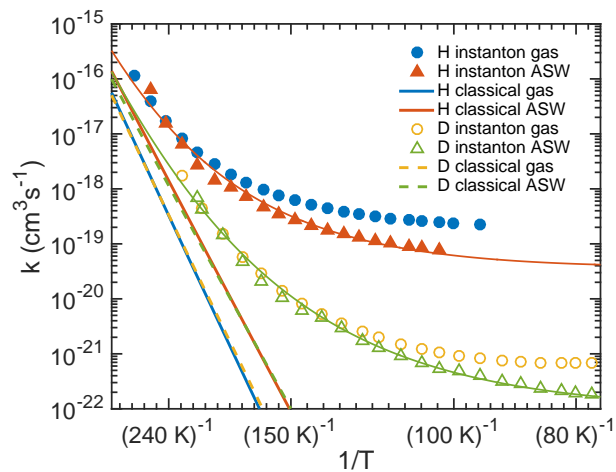}
  \caption{Instanton and classical rate constants for the reactions of H +
    HNCO $\rightarrow$ NH$_2$CO and D + HNCO $\rightarrow$ NHDCO in gas and
    the ER process on the ASW surface. The thin lines represent fits using
    Eq.~(\ref{eq:zt}). }
  \label{fgr:rates}
\end{figure}

\subsection{Tunneling rate constants}

\begin{figure}[h]
\centering
  \includegraphics[height=6.2cm]{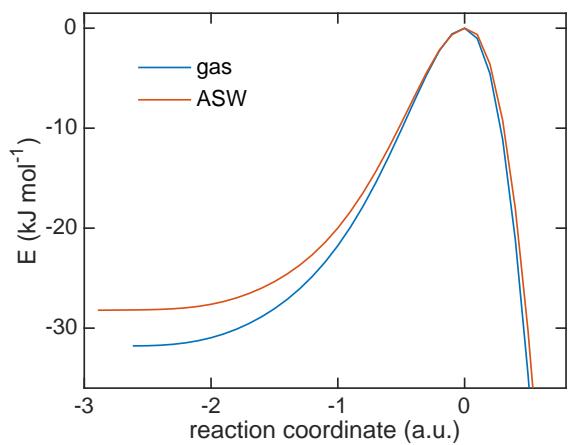}
  \caption{The minimum energy path of the reaction of H + HNCO $\rightarrow$
    NH$_2$CO in the gas phase and on the amorphous solid water surface.}
  \label{fgr:irc}
\end{figure}

Starting from TS1 we calculated rate constants for reaction (\ref{re:nh})
following an ER mechanism on the ASW surface and compared them to the gas
phase reaction treated at the same QM level of theory. The results are shown
in Fig.~\ref{fgr:rates}. The red solid triangles correspond to the rate
constants on the ASW surface, the blue solid circles to the ones of the corresponding 
gas-phase reaction. Instanton rate constant calculations are restricted to temperatures 
below $T_\text{c}$. At high temperature the surface-bound reaction is
slightly faster than the gas-phase reaction; at low temperature the case is
reversed and the gas-phase reaction becomes more efficient. Thus, there is no
significant catalytic effect of the surface. However, the surface of course
still has the effect of dissipating the excess energy of the reaction and
increasing the local concentration of the reactants. Despite the lower
barrier, the tunneling rate constant for the ASW-bound reaction is lower than
the gas phase reaction at low temperature. This demonstrates again that
besides the barrier height, the barrier width is important for the tunneling
efficiency.\cite{kae13} The barrier shapes along the intrinsic reaction
coordinates (IRC) are compared in Fig.~\ref{fgr:irc}, which clearly shows that
the ASW-barrier is lower but broader than the gas-phase barrier which leads to
the lower tunneling rate at low temperature.

Our data allow the comparison between a structural model which contains the
surface explicitly and a gas-phase model for the surface reaction. As
discussed above, the barrier changes only very slightly due to the influence
of the surface and quite independently of the binding site. The resulting
rate constants are very similar. The surface, however, restricts the
rotational motion of the reactant and the transition state. The change in the
rotational partition function is included in the rate constants depicted in
Fig.~\ref{fgr:rates}. One can model a surface by considering only the atoms
HNCO + H explicitly but restricting the rotational motion, i.e. ignoring the
change in the rotational partition function between HNCO and the transition
state. This corresponds to the rotational restriction of both HNCO and the
transition state on the surface. With such an approach, the rate constants
obtained from a gas-phase model are even more similar to those obtained from
the surface model, e.g., at 103~K we find a rate constant on the surface of
$7.8\times 10^{-20}$ cm$^3$~s$^{-1}$, of $8.0\times 10^{-20}$ cm$^3$~s$^{-1}$
for the gas phase model with restricted rotation and of $2.4\times 10^{-19}$
cm$^3$~s$^{-1}$ for the gas phase model with full rotation. For the reaction
under study a gas-phase model with restricted rotation results in
sufficiently accurate surface rate constants.

\begin{table}[htbp]
\small
  \caption{Parameters for rate constants described of the reaction H/D
    + HNCO by  Eq.~(\ref{eq:zt}).\label{tab:fit}}
  \begin{tabular*}{\columnwidth}{@{\extracolsep{\fill}}lrr}
    \hline
   parameter                     & H & D \\
    \hline
    $\alpha$ (cm$^3$ s$^{-1}$)  & $7.22\times 10^{-12}$ & $4.13\times 10^{-12}$\\
    $\beta$ & 1 & 1\\
    $\gamma$ (K) & 2856 & 2887\\
    $T_0$ (K) & 195.8 & 153.4\\
    \hline  
  \end{tabular*}
\end{table}

Rate constants were fitted to Eq.~(\ref{eq:zt}) to facilitate the use
of our results in astrochemical models. The parameters are given in
Table~\ref{tab:fit}, the resulting curves are shown in
Fig.~\ref{fgr:rates} as thin red and green lines. They match the
calculated rate constants reasonably well. We recommend using the fit
in a temperature range close to the range that was used to produce it,
i.e. 1000~K to $\sim$90~K for H+HNCO and 1000~K to $\sim$60~K for
D+HNCO.

The red and blue straight lines in Fig.~\ref{fgr:rates} correspond
to the rate constants neglecting tunneling (but including quantized
vibrations and, thus, the ZPE). Due to the smaller barrier, without tunneling the
surface-bound reaction is always faster than the gas-phase
reaction. Tunneling accelerates the reaction by many orders of
magnitude at low temperature. \clr{Values for the rate constants with
  and without tunneling are
  given in Tables 2 and 3 of the Supporting Information. For example
  at 103~K, tunneling accelerates the gas-phase by a factor of
  $2\times 10^{10}$ and the surface reaction by a factor of
$10^8$. These values increase steeply with decreasing temperature.}

\begin{figure}[h]
\centering
  \includegraphics[height=6.8cm]{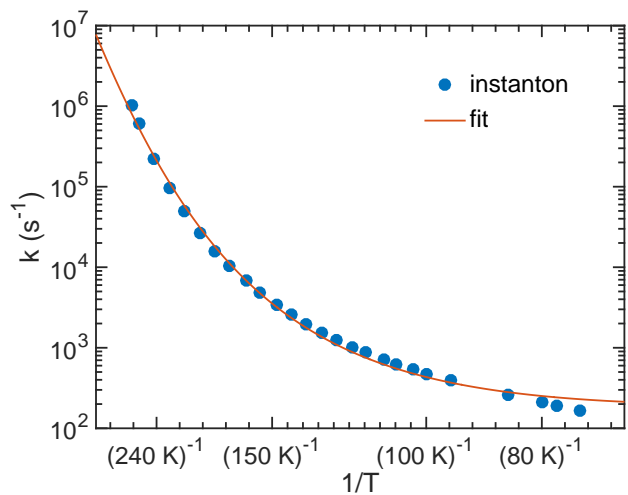}
  \caption{Instanton rate constants for the LH process of reaction
    (\ref{re:nh}) on the ASW surface.}
  \label{fgr:unimol}
\end{figure}

The bimolecular rate constants reported above relate to an ER mechanism. At
low temperature a LH mechanism is more likely. In that case we can assume
HNCO to be stationary on the surface while the H atom diffuses with the
hopping rate constant $k_\text{hop}$ until it meets a HNCO site. Then it can
either react or diffuse away again. The probability for reaction is
$k_\text{react}/(k_\text{react}+k_\text{hop})$ where $k_\text{react}$ is a
unimolecular rate constant which we can calculate. It corresponds to the
process of an encounter complex of H with HNCO reacting to NH$_2$CO. Since H
is bound very weakly on the surface, we were able to optimize such an
encounter complex only for TS2. Its energy is 4.7~kJ~mol$^{-1}$
(1.4~kJ~mol$^{-1}$ with ZPE) below that of the separated reactants. The
resulting rate constants are shown in Fig.~\ref{fgr:unimol}. We fitted the
parameters of Eq.~\ref{eq:zt}, which resulted in $\alpha=3.56\times
10^{10}$~s$^{-1}$, $\gamma= 2503$~K and $T_0=172.9$~K. The parameter $\beta$
was kept to 1 just like in the other fits.

\subsection{Kinetic isotope effects}

In addition, we investigated the kinetic isotope effect (KIE) for
reaction~(\ref{re:nh}). For D + HNCO $\rightarrow$ NHDCO, the crossover
temperature is reduced from 284~K to 218~K on the ASW surface and from 307~K
to 235~K in the gas phase. In Fig.~\ref{fgr:rates} instanton rate constants
for the reactions with deuterium in the gas-phase are shown by yellow circles and
on ASW by green triangles. Similar trends are visible as for the addition of
protium to HNCO, but the rate constants are much smaller. As frequently
observed for tunneling reactions, the KIE increases with decreasing
temperature. At 103~K the KIE for the gas-phase reaction is 231, on the ASW
surface it is 146. Even stronger KIEs can be expected at lower
temperature. The KIEs without tunneling are much smaller as can be seen from
Fig.~\ref{fgr:rates}, which indicates that the KIE is mostly caused by
tunneling rather than by the difference in the ZPE.

\subsection{Alternative gas-phase reaction}

To elucidate a possible role of reaction~(\ref{re:h2co}) for the formation of
\ce{NH2CHO}, we calculated the barrier for the initial reaction channel, the
approach of \ce{NH2} to formaldehyde. We optimized the reactants and the
transition state on the 
M06-2X~\cite{zhao:06}/def2-TZVP~\cite{florian:05} 
level using NWCHEM 6.6~\cite{nwchem} 
and calculated single-point energies \clr{and vibrational frequencies} 
on the UCCSD(T)-F12~\cite{adler:07,knizia:09}/cc-pVTZ-F12~\cite{peterson:08} level. 
The coordinates of the transition structure are given in the supporting
information. In agreement with previous work,\cite{barone:15} we found an
almost submerged barrier on the potential energy surface, +2.7~kJ~mol$^{-1}$ compared
to the separated reactants. Including the ZPE, however,
resulted in a significant barrier of \clr{17.8}~kJ~mol$^{-1}$. \clr{The
crossover temperature is 88.0~K. Thus, tunneling only plays a minor
role above that temperature. The corresponding rate constant for
reaction~(\ref{re:h2co}) at 100~K is $k=1.1\times
10^{-22}$~cm$^3$~s$^{-1}$ if tunneling is neglected and quite a
similar value of $k=5.3\times
10^{-22}$~cm$^3$~s$^{-1}$ if tunneling is approximated via a symmetric
Eckart barrier. Note that above the crossover temperature, instanton
theory is not applicable. These rate constants can only serve as an
upper limit to the full rate constant of reaction~(\ref{re:h2co})
since they only cover the entrance channel. The full reaction contains
additional submerged barriers\cite{barone:15} which might lower the
rate even further. Nevertheless, even these upper bounds are significantly
smaller than the rate constant of $k=2.4\times
10^{-19}$~cm$^3$~s$^{-1}$ for reaction (\ref{re:nh}) at the same
temperature.} Thus, we
conclude that the gas-phase reaction~(\ref{re:h2co}) is expected not to play a
significant role in the formation of \ce{NH2CHO}.

\section{Conclusions}

We investigated binding of HNCO to an ASW surface and subsequent
hydrogenation. Different binding sites with a significant spread of binding
energies were found. The activation barrier for the hydrogenation reaction
turned out to be rather independent of the binding energy.  We calculated the
reaction rate constants for H + HNCO $\rightarrow$ NH$_2$CO in the gas phase
at temperatures of 289~K down to 95~K and on the ASW surface down to 103~K by
combining the QM/MM method with instanton theory.  Although the activation
barrier for the surface reaction is 3.9~kJ~mol$^{-1}$ (4.4~kJ~mol$^{-1}$
including ZPE) lower than in the gas-phase, the ASW surface does not
efficiently accelerate this reaction, but hinders it at temperatures below
240~K.  It demonstrates that the width but not the height of the barrier
dominantly affects the tunneling rate for this system.  In addition, the
deuterated reaction of D + HNCO $\rightarrow$ NHDCO has been investigated both
in the gas-phase and on the ASW surface.  According to the instanton calculations, the
KIEs are 231 and 146 for the gas phase reaction and the surface reaction at
103~K, respectively and expected to be at least similarly strong at even lower
temperature. The strong tunnel effect raises the rate constants to values
which enable hydrogenation of HNCO on the surface of interstellar dust grains,
making this a possible route for the formation of the pre-biotic molecule
formamide. By contrast, the gas-phase route via reaction~(\ref{re:h2co}) seems
inaccessible at low temperature.

\section{Acknowledgments}

This work was financially supported by the European Research Council (ERC)
under the European Union’s Horizon 2020 research and innovation programme
(grant agreement No 646717, TUNNELCHEM). The authors also acknowledge support
for CPU time by the state of Baden-W\"urttemberg through bwHPC and the German
Research Foundation (DFG) through grant no INST 40/467-1 FUGG.

\footnotesize{

\begin{mcitethebibliography}{81}
\providecommand*{\natexlab}[1]{#1}
\providecommand*{\mciteSetBstSublistMode}[1]{}
\providecommand*{\mciteSetBstMaxWidthForm}[2]{}
\providecommand*{\mciteBstWouldAddEndPuncttrue}
  {\def\EndOfBibitem{\unskip.}}
\providecommand*{\mciteBstWouldAddEndPunctfalse}
  {\let\EndOfBibitem\relax}
\providecommand*{\mciteSetBstMidEndSepPunct}[3]{}
\providecommand*{\mciteSetBstSublistLabelBeginEnd}[3]{}
\providecommand*{\EndOfBibitem}{}
\mciteSetBstSublistMode{f}
\mciteSetBstMaxWidthForm{subitem}
{(\emph{\alph{mcitesubitemcount}})}
\mciteSetBstSublistLabelBeginEnd{\mcitemaxwidthsubitemform\space}
{\relax}{\relax}

\bibitem[{Rubin} \emph{et~al.}(1971){Rubin}, {Swenson}, {Benson}, {Tigelaar},
  and {Flygare}]{rubin:71}
R.~H. {Rubin}, G.~W. {Swenson}, Jr., R.~C. {Benson}, H.~L. {Tigelaar} and W.~H.
  {Flygare}, \emph{{A}strophys {L}ett}, 1971, \textbf{169}, L39\relax
\mciteBstWouldAddEndPuncttrue
\mciteSetBstMidEndSepPunct{\mcitedefaultmidpunct}
{\mcitedefaultendpunct}{\mcitedefaultseppunct}\relax
\EndOfBibitem
\bibitem[{Bisschop} \emph{et~al.}(2007){Bisschop}, {J{\o}rgensen}, {van
  Dishoeck}, and {de Wachter}]{bisschop:07}
S.~E. {Bisschop}, J.~K. {J{\o}rgensen}, E.~F. {van Dishoeck} and E.~B.~M. {de
  Wachter}, \emph{Astron. Astrophys.}, 2007, \textbf{465}, 913--929\relax
\mciteBstWouldAddEndPuncttrue
\mciteSetBstMidEndSepPunct{\mcitedefaultmidpunct}
{\mcitedefaultendpunct}{\mcitedefaultseppunct}\relax
\EndOfBibitem
\bibitem[{Yamaguchi} \emph{et~al.}(2012){Yamaguchi}, {Takano}, {Watanabe},
  {Sakai}, {Sakai}, {Liu}, {Su}, {Hirano}, {Takakuwa}, {Aikawa}, {Nomura}, and
  {Yamamoto}]{yama:12}
T.~{Yamaguchi}, S.~{Takano}, Y.~{Watanabe}, N.~{Sakai}, T.~{Sakai}, S.-Y.
  {Liu}, Y.-N. {Su}, N.~{Hirano}, S.~{Takakuwa}, Y.~{Aikawa}, H.~{Nomura} and
  S.~{Yamamoto}, \emph{Publ. Astron. Soc. Japan}, 2012, \textbf{64}, 105\relax
\mciteBstWouldAddEndPuncttrue
\mciteSetBstMidEndSepPunct{\mcitedefaultmidpunct}
{\mcitedefaultendpunct}{\mcitedefaultseppunct}\relax
\EndOfBibitem
\bibitem[{Mendoza} \emph{et~al.}(2014){Mendoza}, {Lefloch},
  {L{\'o}pez-Sepulcre}, {Ceccarelli}, {Codella}, {Boechat-Roberty}, and
  {Bachiller}]{mendoza:14}
E.~{Mendoza}, B.~{Lefloch}, A.~{L{\'o}pez-Sepulcre}, C.~{Ceccarelli},
  C.~{Codella}, H.~M. {Boechat-Roberty} and R.~{Bachiller}, \emph{{Mon. Not. R.
  Astron. Soc.}}, 2014, \textbf{445}, 151--161\relax
\mciteBstWouldAddEndPuncttrue
\mciteSetBstMidEndSepPunct{\mcitedefaultmidpunct}
{\mcitedefaultendpunct}{\mcitedefaultseppunct}\relax
\EndOfBibitem
\bibitem[{Bockel{\'e}e-Morvan} \emph{et~al.}(2000){Bockel{\'e}e-Morvan}, {Lis},
  {Wink}, {Despois}, {Crovisier}, {Bachiller}, {Benford}, {Biver}, {Colom},
  {Davies}, {G{\'e}rard}, {Germain}, {Houde}, {Mehringer}, {Moreno}, {Paubert},
  {Phillips}, and {Rauer}]{bockelee:00}
D.~{Bockel{\'e}e-Morvan}, D.~C. {Lis}, J.~E. {Wink}, D.~{Despois},
  J.~{Crovisier}, R.~{Bachiller}, D.~J. {Benford}, N.~{Biver}, P.~{Colom},
  J.~K. {Davies}, E.~{G{\'e}rard}, B.~{Germain}, M.~{Houde}, D.~{Mehringer},
  R.~{Moreno}, G.~{Paubert}, T.~G. {Phillips} and H.~{Rauer}, \emph{Astron.
  Astrophys.}, 2000, \textbf{353}, 1101--1114\relax
\mciteBstWouldAddEndPuncttrue
\mciteSetBstMidEndSepPunct{\mcitedefaultmidpunct}
{\mcitedefaultendpunct}{\mcitedefaultseppunct}\relax
\EndOfBibitem
\bibitem[{L{\'o}pez-Sepulcre} \emph{et~al.}(2015){L{\'o}pez-Sepulcre}, {Jaber},
  {Mendoza}, {Lefloch}, {Ceccarelli}, {Vastel}, {Bachiller}, {Cernicharo},
  {Codella}, {Kahane}, {Kama}, and {Tafalla}]{lopez:15b}
A.~{L{\'o}pez-Sepulcre}, A.~A. {Jaber}, E.~{Mendoza}, B.~{Lefloch},
  C.~{Ceccarelli}, C.~{Vastel}, R.~{Bachiller}, J.~{Cernicharo}, C.~{Codella},
  C.~{Kahane}, M.~{Kama} and M.~{Tafalla}, \emph{Mon. Not. R. Astron. Soc.},
  2015, \textbf{449}, 2438--2458\relax
\mciteBstWouldAddEndPuncttrue
\mciteSetBstMidEndSepPunct{\mcitedefaultmidpunct}
{\mcitedefaultendpunct}{\mcitedefaultseppunct}\relax
\EndOfBibitem
\bibitem[{Raunier} \emph{et~al.}(2004){Raunier}, {Chiavassa}, {Duvernay},
  {Borget}, {Aycard}, {Dartois}, and {d'Hendecourt}]{raunier:04}
S.~{Raunier}, T.~{Chiavassa}, F.~{Duvernay}, F.~{Borget}, J.~P. {Aycard},
  E.~{Dartois} and L.~{d'Hendecourt}, \emph{Astron. Astrophys.}, 2004,
  \textbf{416}, 165--169\relax
\mciteBstWouldAddEndPuncttrue
\mciteSetBstMidEndSepPunct{\mcitedefaultmidpunct}
{\mcitedefaultendpunct}{\mcitedefaultseppunct}\relax
\EndOfBibitem
\bibitem[{Garrod} \emph{et~al.}(2008){Garrod}, {Widicus Weaver}, and
  {Herbst}]{garrod:08}
R.~T. {Garrod}, S.~L. {Widicus Weaver} and E.~{Herbst}, \emph{Astrophys. J.},
  2008, \textbf{682}, 283--302\relax
\mciteBstWouldAddEndPuncttrue
\mciteSetBstMidEndSepPunct{\mcitedefaultmidpunct}
{\mcitedefaultendpunct}{\mcitedefaultseppunct}\relax
\EndOfBibitem
\bibitem[{Redondo} \emph{et~al.}(2014){Redondo}, {Barrientos}, and
  {Largo}]{redondo:14}
P.~{Redondo}, C.~{Barrientos} and A.~{Largo}, \emph{Astrophys. J.}, 2014,
  \textbf{793}, 32--39\relax
\mciteBstWouldAddEndPuncttrue
\mciteSetBstMidEndSepPunct{\mcitedefaultmidpunct}
{\mcitedefaultendpunct}{\mcitedefaultseppunct}\relax
\EndOfBibitem
\bibitem[{Nguyen} \emph{et~al.}(1996){Nguyen}, {Sengupta}, {Vereecken},
  {Peeters}, , and {Vanquickenborne}]{nguyen:96}
M.~T. {Nguyen}, D.~{Sengupta}, L.~{Vereecken}, J.~{Peeters},  and L.~G.
  {Vanquickenborne}, \emph{J. Phys. Chem.}, 1996, \textbf{100},
  1615--1621\relax
\mciteBstWouldAddEndPuncttrue
\mciteSetBstMidEndSepPunct{\mcitedefaultmidpunct}
{\mcitedefaultendpunct}{\mcitedefaultseppunct}\relax
\EndOfBibitem
\bibitem[{Noble} \emph{et~al.}(2015){Noble}, {Theule}, {Congiu}, {Dulieu},
  {Bonnin}, {Bassas}, {Duvernay}, {Danger}, and {Chiavassa}]{noble:15}
J.~A. {Noble}, P.~{Theule}, E.~{Congiu}, F.~{Dulieu}, M.~{Bonnin}, A.~{Bassas},
  F.~{Duvernay}, G.~{Danger} and T.~{Chiavassa}, \emph{Astron. Astrophys.},
  2015, \textbf{576}, A91\relax
\mciteBstWouldAddEndPuncttrue
\mciteSetBstMidEndSepPunct{\mcitedefaultmidpunct}
{\mcitedefaultendpunct}{\mcitedefaultseppunct}\relax
\EndOfBibitem
\bibitem[{Barone} \emph{et~al.}(2015){Barone}, {Latouche}, {Skouteris},
  {Vazart}, {Balucani}, {Ceccarelli}, and {Lefloch}]{barone:15}
V.~{Barone}, C.~{Latouche}, D.~{Skouteris}, F.~{Vazart}, N.~{Balucani},
  C.~{Ceccarelli} and B.~{Lefloch}, \emph{{Astron. Astrophys.}}, 2015,
  \textbf{453}, L31--L35\relax
\mciteBstWouldAddEndPuncttrue
\mciteSetBstMidEndSepPunct{\mcitedefaultmidpunct}
{\mcitedefaultendpunct}{\mcitedefaultseppunct}\relax
\EndOfBibitem
\bibitem[Hama and Watanabe(2013)]{hama:13}
T.~Hama and N.~Watanabe, \emph{Chem. Rev.}, 2013, \textbf{113},
  8783--8839\relax
\mciteBstWouldAddEndPuncttrue
\mciteSetBstMidEndSepPunct{\mcitedefaultmidpunct}
{\mcitedefaultendpunct}{\mcitedefaultseppunct}\relax
\EndOfBibitem
\bibitem[Meisner and K\"astner(2016)]{mei16}
J.~Meisner and J.~K\"astner, \emph{Angew. Chem. Int. Ed.}, 2016, \textbf{55},
  5400--5413\relax
\mciteBstWouldAddEndPuncttrue
\mciteSetBstMidEndSepPunct{\mcitedefaultmidpunct}
{\mcitedefaultendpunct}{\mcitedefaultseppunct}\relax
\EndOfBibitem
\bibitem[Phillips \emph{et~al.}(2005)Phillips, Braun, Wang, Gumbart,
  Tajkhorshid, Villa, Chipot, Skeel, Kal\'e, and Schulten]{phillips:05}
J.~C. Phillips, R.~Braun, W.~Wang, J.~Gumbart, E.~Tajkhorshid, E.~Villa,
  C.~Chipot, R.~D. Skeel, L.~Kal\'e and K.~Schulten, \emph{J. Comput. Chem.},
  2005, \textbf{26}, 1781\relax
\mciteBstWouldAddEndPuncttrue
\mciteSetBstMidEndSepPunct{\mcitedefaultmidpunct}
{\mcitedefaultendpunct}{\mcitedefaultseppunct}\relax
\EndOfBibitem
\bibitem[Humphrey \emph{et~al.}(1996)Humphrey, Dalke, and Schulten]{hum96a}
W.~Humphrey, A.~Dalke and K.~Schulten, \emph{J. Molec. Graphics}, 1996,
  \textbf{14}, 33\relax
\mciteBstWouldAddEndPuncttrue
\mciteSetBstMidEndSepPunct{\mcitedefaultmidpunct}
{\mcitedefaultendpunct}{\mcitedefaultseppunct}\relax
\EndOfBibitem
\bibitem[Jorgensen \emph{et~al.}(1983)Jorgensen, Chandrasekhar, Madura, Impey,
  and Klein]{jor83}
W.~L. Jorgensen, J.~Chandrasekhar, J.~D. Madura, R.~W. Impey and M.~L. Klein,
  \emph{J. Chem. Phys.}, 1983, \textbf{79}, 926\relax
\mciteBstWouldAddEndPuncttrue
\mciteSetBstMidEndSepPunct{\mcitedefaultmidpunct}
{\mcitedefaultendpunct}{\mcitedefaultseppunct}\relax
\EndOfBibitem
\bibitem[Warshel and Karplus(1972)]{warshel:72}
A.~Warshel and M.~Karplus, \emph{J. Am. Chem. Soc}, 1972, \textbf{94},
  5612--5625\relax
\mciteBstWouldAddEndPuncttrue
\mciteSetBstMidEndSepPunct{\mcitedefaultmidpunct}
{\mcitedefaultendpunct}{\mcitedefaultseppunct}\relax
\EndOfBibitem
\bibitem[Warshel and Levitt(1976)]{warshel:76}
A.~Warshel and M.~Levitt, \emph{J. Mol. biol.}, 1976, \textbf{103}, 
227--249\relax
\mciteBstWouldAddEndPuncttrue
\mciteSetBstMidEndSepPunct{\mcitedefaultmidpunct}
{\mcitedefaultendpunct}{\mcitedefaultseppunct}\relax
\EndOfBibitem
\bibitem[Sherwood \emph{et~al.}(2003)Sherwood, de~Vries, Guest, Schreckenbach,
  Catlow, French, Sokol, Bromley, Thiel, Turner, Billeter, Terstegen, Thiel,
  Kendrick, Rogers, Casci, Watson, King, Karlsen, Sj{\o}voll, Fahmi,
  Sch{\"a}fer, and Lennartz]{she03}
P.~Sherwood, A.~H. de~Vries, M.~F. Guest, G.~Schreckenbach, C.~R.~A. Catlow,
  S.~A. French, A.~A. Sokol, S.~T. Bromley, W.~Thiel, A.~J. Turner,
  S.~Billeter, F.~Terstegen, S.~Thiel, J.~Kendrick, S.~C. Rogers, J.~Casci,
  M.~Watson, F.~King, E.~Karlsen, M.~Sj{\o}voll, A.~Fahmi, A.~Sch{\"a}fer and
  C.~Lennartz, \emph{J. Mol. Struct. (THEOCHEM)}, 2003, \textbf{632}, 1\relax
\mciteBstWouldAddEndPuncttrue
\mciteSetBstMidEndSepPunct{\mcitedefaultmidpunct}
{\mcitedefaultendpunct}{\mcitedefaultseppunct}\relax
\EndOfBibitem
\bibitem[Metz \emph{et~al.}(2014)Metz, K\"astner, Sokol, Keal, and
  Sherwood]{met14}
S.~Metz, J.~K\"astner, A.~A. Sokol, T.~W. Keal and P.~Sherwood, \emph{WIREs
  Comput. Mol. Sci.}, 2014, \textbf{4}, 101\relax
\mciteBstWouldAddEndPuncttrue
\mciteSetBstMidEndSepPunct{\mcitedefaultmidpunct}
{\mcitedefaultendpunct}{\mcitedefaultseppunct}\relax
\EndOfBibitem
\bibitem[Stephens \emph{et~al.}(1994)Stephens, Devlin, Chabalowski, and
  Frisch]{ste94}
P.~Stephens, F.~Devlin, C.~Chabalowski and M.~Frisch, \emph{J. Phys. Chem.},
  1994, \textbf{98}, 11623\relax
\mciteBstWouldAddEndPuncttrue
\mciteSetBstMidEndSepPunct{\mcitedefaultmidpunct}
{\mcitedefaultendpunct}{\mcitedefaultseppunct}\relax
\EndOfBibitem
\bibitem[Rappoport and Furche(2010)]{rappoport:10}
D.~Rappoport and F.~Furche, \emph{J. Chem. Phys.}, 2010, \textbf{133},
  134105\relax
\mciteBstWouldAddEndPuncttrue
\mciteSetBstMidEndSepPunct{\mcitedefaultmidpunct}
{\mcitedefaultendpunct}{\mcitedefaultseppunct}\relax
\EndOfBibitem
\bibitem[Becke(1993)]{becke:93b}
A.~D. Becke, \emph{J. Chem. Phys.}, 1993, \textbf{98}, 1372--1377\relax
\mciteBstWouldAddEndPuncttrue
\mciteSetBstMidEndSepPunct{\mcitedefaultmidpunct}
{\mcitedefaultendpunct}{\mcitedefaultseppunct}\relax
\EndOfBibitem
\bibitem[Lee \emph{et~al.}(1988)Lee, Yang, and Parr]{lee:88}
C.~Lee, W.~Yang and R.~G. Parr, \emph{Phys. Rev. B}, 1988, \textbf{37},
  785--789\relax
\mciteBstWouldAddEndPuncttrue
\mciteSetBstMidEndSepPunct{\mcitedefaultmidpunct}
{\mcitedefaultendpunct}{\mcitedefaultseppunct}\relax
\EndOfBibitem
\bibitem[Grimme \emph{et~al.}(2010)Grimme, Antony, Ehrlich, and
  Krieg]{grimme:10}
S.~Grimme, J.~Antony, S.~Ehrlich and H.~Krieg, \emph{J. Chem. Phys.}, 2010,
  \textbf{132}, 154104\relax
\mciteBstWouldAddEndPuncttrue
\mciteSetBstMidEndSepPunct{\mcitedefaultmidpunct}
{\mcitedefaultendpunct}{\mcitedefaultseppunct}\relax
\EndOfBibitem
\bibitem[Weigend and Ahlrichs(2005)]{florian:05}
F.~Weigend and R.~Ahlrichs, \emph{Phys. Chem. Chem. Phys.}, 2005, \textbf{7},
  3297--3305\relax
\mciteBstWouldAddEndPuncttrue
\mciteSetBstMidEndSepPunct{\mcitedefaultmidpunct}
{\mcitedefaultendpunct}{\mcitedefaultseppunct}\relax
\EndOfBibitem
\bibitem[tur()]{turbomole}
\emph{{TURBOMOLE V}, a development of {University of Karlsruhe} and
  {Forschungszentrum Karlsruhe GmbH}, 1989-2007, {TURBOMOLE GmbH}, since 2007;
  available from {\tt http://www.turbomole.com}.}\relax
\mciteBstWouldAddEndPunctfalse
\mciteSetBstMidEndSepPunct{\mcitedefaultmidpunct}
{}{\mcitedefaultseppunct}\relax
\EndOfBibitem
\bibitem[Smith \emph{et~al.}(2002)Smith, Yong, and Rodger]{smith:02}
W.~Smith, C.~Yong and P.~Rodger, \emph{Mol. Sim.}, 2002, \textbf{28},
  385--471\relax
\mciteBstWouldAddEndPuncttrue
\mciteSetBstMidEndSepPunct{\mcitedefaultmidpunct}
{\mcitedefaultendpunct}{\mcitedefaultseppunct}\relax
\EndOfBibitem
\bibitem[MacKerell~Jr. \emph{et~al.}(1998)MacKerell~Jr., Bashford, Bellott,
  Dunbrack~Jr., Evanseck, Field, Fischer, Gao, Guo, Ha, Joseph-McCarthy,
  Kuchnir, Kuczera, Lau, Mattos, Michnick, Ngo, Nguyen, Prodhom, Reiher~III,
  Roux, Schlenkrich, Smith, Stote, Straub, Watanabe, Wiorkiewicz-Kuczera, Yin,
  and Karplus]{mackerell:98}
A.~D. MacKerell~Jr., D.~Bashford, R.~L. Bellott, R.~L. Dunbrack~Jr., J.~D.
  Evanseck, M.~J. Field, S.~Fischer, J.~Gao, H.~Guo, S.~Ha, D.~Joseph-McCarthy,
  L.~Kuchnir, K.~Kuczera, F.~T.~K. Lau, C.~Mattos, S.~Michnick, T.~Ngo, D.~T.
  Nguyen, B.~Prodhom, W.~E. Reiher~III, B.~Roux, M.~Schlenkrich, J.~C. Smith,
  R.~Stote, J.~Straub, M.~Watanabe, J.~Wiorkiewicz-Kuczera, D.~Yin and
  M.~Karplus, \emph{J. Phys. Chem. B}, 1998, \textbf{102}, 3586\relax
\mciteBstWouldAddEndPuncttrue
\mciteSetBstMidEndSepPunct{\mcitedefaultmidpunct}
{\mcitedefaultendpunct}{\mcitedefaultseppunct}\relax
\EndOfBibitem
\bibitem[MacKerell~Jr. and Banavali(2000)]{mackerell:00}
A.~MacKerell~Jr. and N.~K. Banavali, \emph{J. Comput. Chem.}, 2000,
  \textbf{21}, 105\relax
\mciteBstWouldAddEndPuncttrue
\mciteSetBstMidEndSepPunct{\mcitedefaultmidpunct}
{\mcitedefaultendpunct}{\mcitedefaultseppunct}\relax
\EndOfBibitem
\bibitem[MacKerell~Jr. \emph{et~al.}(2004)MacKerell~Jr., Feig, and
  Brooks~III]{mackerell:04}
A.~D. MacKerell~Jr., M.~Feig and C.~Brooks~III, \emph{J. Comput. Chem.}, 2004,
  \textbf{25}, 1400\relax
\mciteBstWouldAddEndPuncttrue
\mciteSetBstMidEndSepPunct{\mcitedefaultmidpunct}
{\mcitedefaultendpunct}{\mcitedefaultseppunct}\relax
\EndOfBibitem
\bibitem[K\"astner \emph{et~al.}(2009)K\"astner, Carr, Keal, Thiel, Wander, and
  Sherwood]{kaestner:09}
J.~K\"astner, J.~M. Carr, T.~W. Keal, W.~Thiel, A.~Wander and P.~Sherwood,
  \emph{J. Phys. Chem. A}, 2009, \textbf{113}, 11856\relax
\mciteBstWouldAddEndPuncttrue
\mciteSetBstMidEndSepPunct{\mcitedefaultmidpunct}
{\mcitedefaultendpunct}{\mcitedefaultseppunct}\relax
\EndOfBibitem
\bibitem[Henkelman and J{\'o}nsson(1999)]{hen99}
G.~Henkelman and H.~J{\'o}nsson, \emph{J. Chem. Phys.}, 1999, \textbf{111},
  7010\relax
\mciteBstWouldAddEndPuncttrue
\mciteSetBstMidEndSepPunct{\mcitedefaultmidpunct}
{\mcitedefaultendpunct}{\mcitedefaultseppunct}\relax
\EndOfBibitem
\bibitem[Heyden \emph{et~al.}(2005)Heyden, Bell, and Keil]{hey05}
A.~Heyden, A.~T. Bell and F.~J. Keil, \emph{J. Chem. Phys.}, 2005,
  \textbf{123}, 224101\relax
\mciteBstWouldAddEndPuncttrue
\mciteSetBstMidEndSepPunct{\mcitedefaultmidpunct}
{\mcitedefaultendpunct}{\mcitedefaultseppunct}\relax
\EndOfBibitem
\bibitem[K\"astner and Sherwood(2008)]{kae08}
J.~K\"astner and P.~Sherwood, \emph{J. Chem. Phys.}, 2008, \textbf{128},
  014106\relax
\mciteBstWouldAddEndPuncttrue
\mciteSetBstMidEndSepPunct{\mcitedefaultmidpunct}
{\mcitedefaultendpunct}{\mcitedefaultseppunct}\relax
\EndOfBibitem
\bibitem[Rommel \emph{et~al.}(2011)Rommel, Goumans, and K\"astner]{rom11}
J.~B. Rommel, T.~P.~M. Goumans and J.~K\"astner, \emph{J. Chem. Theory
  Comput.}, 2011, \textbf{7}, 690\relax
\mciteBstWouldAddEndPuncttrue
\mciteSetBstMidEndSepPunct{\mcitedefaultmidpunct}
{\mcitedefaultendpunct}{\mcitedefaultseppunct}\relax
\EndOfBibitem
\bibitem[Rommel and K\"astner(2011)]{rom11b}
J.~B. Rommel and J.~K\"astner, \emph{J. Chem. Phys.}, 2011, \textbf{134},
  184107\relax
\mciteBstWouldAddEndPuncttrue
\mciteSetBstMidEndSepPunct{\mcitedefaultmidpunct}
{\mcitedefaultendpunct}{\mcitedefaultseppunct}\relax
\EndOfBibitem
\bibitem[Langer(1967)]{lan67}
J.~S. Langer, \emph{Ann. Phys. (N.Y.)}, 1967, \textbf{41}, 108\relax
\mciteBstWouldAddEndPuncttrue
\mciteSetBstMidEndSepPunct{\mcitedefaultmidpunct}
{\mcitedefaultendpunct}{\mcitedefaultseppunct}\relax
\EndOfBibitem
\bibitem[Miller(1975)]{mil75}
W.~H. Miller, \emph{J. Chem. Phys.}, 1975, \textbf{62}, 1899\relax
\mciteBstWouldAddEndPuncttrue
\mciteSetBstMidEndSepPunct{\mcitedefaultmidpunct}
{\mcitedefaultendpunct}{\mcitedefaultseppunct}\relax
\EndOfBibitem
\bibitem[Coleman(1977)]{col77}
S.~Coleman, \emph{Phys. Rev. D}, 1977, \textbf{15}, 2929\relax
\mciteBstWouldAddEndPuncttrue
\mciteSetBstMidEndSepPunct{\mcitedefaultmidpunct}
{\mcitedefaultendpunct}{\mcitedefaultseppunct}\relax
\EndOfBibitem
\bibitem[Callan~Jr. and Coleman(1977)]{cal77}
C.~G. Callan~Jr. and S.~Coleman, \emph{Phys. Rev. D}, 1977, \textbf{16},
  1762\relax
\mciteBstWouldAddEndPuncttrue
\mciteSetBstMidEndSepPunct{\mcitedefaultmidpunct}
{\mcitedefaultendpunct}{\mcitedefaultseppunct}\relax
\EndOfBibitem
\bibitem[Althorpe(2011)]{alt11}
S.~C. Althorpe, \emph{J. Chem. Phys.}, 2011, \textbf{134}, 114104\relax
\mciteBstWouldAddEndPuncttrue
\mciteSetBstMidEndSepPunct{\mcitedefaultmidpunct}
{\mcitedefaultendpunct}{\mcitedefaultseppunct}\relax
\EndOfBibitem
\bibitem[Richardson(2016)]{ric16}
J.~O. Richardson, \emph{J. Chem. Phys.}, 2016, \textbf{144}, 114106\relax
\mciteBstWouldAddEndPuncttrue
\mciteSetBstMidEndSepPunct{\mcitedefaultmidpunct}
{\mcitedefaultendpunct}{\mcitedefaultseppunct}\relax
\EndOfBibitem
\bibitem[Affleck(1981)]{aff81}
I.~Affleck, \emph{Phys. Rev. Lett.}, 1981, \textbf{46}, 388--391\relax
\mciteBstWouldAddEndPuncttrue
\mciteSetBstMidEndSepPunct{\mcitedefaultmidpunct}
{\mcitedefaultendpunct}{\mcitedefaultseppunct}\relax
\EndOfBibitem
\bibitem[Coleman(1988)]{col88}
S.~Coleman, \emph{Nucl. Phys. B}, 1988, \textbf{298}, 178\relax
\mciteBstWouldAddEndPuncttrue
\mciteSetBstMidEndSepPunct{\mcitedefaultmidpunct}
{\mcitedefaultendpunct}{\mcitedefaultseppunct}\relax
\EndOfBibitem
\bibitem[H\"anggi \emph{et~al.}(1990)H\"anggi, Talkner, and Borkovec]{han90}
P.~H\"anggi, P.~Talkner and M.~Borkovec, \emph{Rev. Mod. Phys.}, 1990,
  \textbf{62}, 251\relax
\mciteBstWouldAddEndPuncttrue
\mciteSetBstMidEndSepPunct{\mcitedefaultmidpunct}
{\mcitedefaultendpunct}{\mcitedefaultseppunct}\relax
\EndOfBibitem
\bibitem[Benderskii \emph{et~al.}(1994)Benderskii, Makarov, and Wight]{ben94}
V.~A. Benderskii, D.~E. Makarov and C.~A. Wight, \emph{Adv. Chem. Phys.}, 1994,
  \textbf{88}, 55\relax
\mciteBstWouldAddEndPuncttrue
\mciteSetBstMidEndSepPunct{\mcitedefaultmidpunct}
{\mcitedefaultendpunct}{\mcitedefaultseppunct}\relax
\EndOfBibitem
\bibitem[Messina \emph{et~al.}(1995)Messina, Schenter, and Garrett]{mes95}
M.~Messina, G.~K. Schenter and B.~C. Garrett, \emph{J. Chem. Phys.}, 1995,
  \textbf{103}, 3430\relax
\mciteBstWouldAddEndPuncttrue
\mciteSetBstMidEndSepPunct{\mcitedefaultmidpunct}
{\mcitedefaultendpunct}{\mcitedefaultseppunct}\relax
\EndOfBibitem
\bibitem[Richardson and Althorpe(2009)]{ric09}
J.~O. Richardson and S.~C. Althorpe, \emph{J. Chem. Phys.}, 2009, \textbf{131},
  214106\relax
\mciteBstWouldAddEndPuncttrue
\mciteSetBstMidEndSepPunct{\mcitedefaultmidpunct}
{\mcitedefaultendpunct}{\mcitedefaultseppunct}\relax
\EndOfBibitem
\bibitem[Zhang \emph{et~al.}(2014)Zhang, Rommel, Cvita{\v s}, and
  Althorpe]{zha14}
Y.~Zhang, J.~B. Rommel, M.~T. Cvita{\v s} and S.~C. Althorpe, \emph{Phys. Chem.
  Chem. Phys.}, 2014, \textbf{16}, 24292--24300\relax
\mciteBstWouldAddEndPuncttrue
\mciteSetBstMidEndSepPunct{\mcitedefaultmidpunct}
{\mcitedefaultendpunct}{\mcitedefaultseppunct}\relax
\EndOfBibitem
\bibitem[Gillan(1987)]{gillan:87}
M.~J. Gillan, \emph{J. Phys. C}, 1987, \textbf{20}, 3621\relax
\mciteBstWouldAddEndPuncttrue
\mciteSetBstMidEndSepPunct{\mcitedefaultmidpunct}
{\mcitedefaultendpunct}{\mcitedefaultseppunct}\relax
\EndOfBibitem
\bibitem[Chapman \emph{et~al.}(1975)Chapman, Garrett, and Miller]{cha75}
S.~Chapman, B.~C. Garrett and W.~H. Miller, \emph{J. Chem. Phys.}, 1975,
  \textbf{63}, 2710\relax
\mciteBstWouldAddEndPuncttrue
\mciteSetBstMidEndSepPunct{\mcitedefaultmidpunct}
{\mcitedefaultendpunct}{\mcitedefaultseppunct}\relax
\EndOfBibitem
\bibitem[Mills and J{\'o}nsson(1994)]{mil94}
G.~Mills and H.~J{\'o}nsson, \emph{Phys. Rev. Lett.}, 1994, \textbf{72},
  1124\relax
\mciteBstWouldAddEndPuncttrue
\mciteSetBstMidEndSepPunct{\mcitedefaultmidpunct}
{\mcitedefaultendpunct}{\mcitedefaultseppunct}\relax
\EndOfBibitem
\bibitem[Mills \emph{et~al.}(1995)Mills, J{\'o}nsson, and Schenter]{mil95}
G.~Mills, H.~J{\'o}nsson and G.~K. Schenter, \emph{Surf. Sci.}, 1995,
  \textbf{324}, 305--337\relax
\mciteBstWouldAddEndPuncttrue
\mciteSetBstMidEndSepPunct{\mcitedefaultmidpunct}
{\mcitedefaultendpunct}{\mcitedefaultseppunct}\relax
\EndOfBibitem
\bibitem[Mills \emph{et~al.}(1997)Mills, Schenter, Makarov, and
  J{\'o}nsson]{mil97}
G.~Mills, G.~K. Schenter, D.~E. Makarov and H.~J{\'o}nsson, \emph{Chem. Phys.
  Lett.}, 1997, \textbf{278}, 91\relax
\mciteBstWouldAddEndPuncttrue
\mciteSetBstMidEndSepPunct{\mcitedefaultmidpunct}
{\mcitedefaultendpunct}{\mcitedefaultseppunct}\relax
\EndOfBibitem
\bibitem[Siebrand \emph{et~al.}(1999)Siebrand, Smedarchina, Zgierski, and
  Fern\'andez-Ramos]{sie99}
W.~Siebrand, Z.~Smedarchina, M.~Z. Zgierski and A.~Fern\'andez-Ramos,
  \emph{Int. Rev. Phys. Chem.}, 1999, \textbf{18}, 5\relax
\mciteBstWouldAddEndPuncttrue
\mciteSetBstMidEndSepPunct{\mcitedefaultmidpunct}
{\mcitedefaultendpunct}{\mcitedefaultseppunct}\relax
\EndOfBibitem
\bibitem[Smedarchina \emph{et~al.}(2003)Smedarchina, Siebrand,
  Fern\'andez-Ramos, and Cui]{sme03}
Z.~Smedarchina, W.~Siebrand, A.~Fern\'andez-Ramos and Q.~Cui, \emph{J. Am.
  Chem. Soc.}, 2003, \textbf{125}, 243--251\relax
\mciteBstWouldAddEndPuncttrue
\mciteSetBstMidEndSepPunct{\mcitedefaultmidpunct}
{\mcitedefaultendpunct}{\mcitedefaultseppunct}\relax
\EndOfBibitem
\bibitem[Qian \emph{et~al.}(2007)Qian, Ren, Shi, E, and Sheng]{qia07}
T.~Qian, W.~Ren, J.~Shi, W.~E and P.~Sheng, \emph{Physica A}, 2007,
  \textbf{379}, 491\relax
\mciteBstWouldAddEndPuncttrue
\mciteSetBstMidEndSepPunct{\mcitedefaultmidpunct}
{\mcitedefaultendpunct}{\mcitedefaultseppunct}\relax
\EndOfBibitem
\bibitem[Andersson \emph{et~al.}(2009)Andersson, Nyman, Arnaldsson, Manthe, and
  J{\'o}nsson]{and09}
S.~Andersson, G.~Nyman, A.~Arnaldsson, U.~Manthe and H.~J{\'o}nsson, \emph{J.
  Phys. Chem. A}, 2009, \textbf{113}, 4468\relax
\mciteBstWouldAddEndPuncttrue
\mciteSetBstMidEndSepPunct{\mcitedefaultmidpunct}
{\mcitedefaultendpunct}{\mcitedefaultseppunct}\relax
\EndOfBibitem
\bibitem[Goumans and Andersson(2010)]{gou10a}
T.~P.~M. Goumans and S.~Andersson, \emph{Mon. Not. R. Astron. Soc.}, 2010,
  \textbf{406}, 2213--2217\relax
\mciteBstWouldAddEndPuncttrue
\mciteSetBstMidEndSepPunct{\mcitedefaultmidpunct}
{\mcitedefaultendpunct}{\mcitedefaultseppunct}\relax
\EndOfBibitem
\bibitem[Goumans(2011)]{gou11}
T.~P.~M. Goumans, \emph{Mon. Not. R. Astron. Soc.}, 2011, \textbf{415},
  3129--3134\relax
\mciteBstWouldAddEndPuncttrue
\mciteSetBstMidEndSepPunct{\mcitedefaultmidpunct}
{\mcitedefaultendpunct}{\mcitedefaultseppunct}\relax
\EndOfBibitem
\bibitem[Goumans(2011)]{gou11b}
T.~P.~M. Goumans, \emph{Mon. Not. R. Astron. Soc.}, 2011, \textbf{413},
  2615--2620\relax
\mciteBstWouldAddEndPuncttrue
\mciteSetBstMidEndSepPunct{\mcitedefaultmidpunct}
{\mcitedefaultendpunct}{\mcitedefaultseppunct}\relax
\EndOfBibitem
\bibitem[Goumans and K\"astner(2010)]{gou10}
T.~P.~M. Goumans and J.~K\"astner, \emph{Angew. Chem. Int. Ed.}, 2010,
  \textbf{49}, 7350--7352\relax
\mciteBstWouldAddEndPuncttrue
\mciteSetBstMidEndSepPunct{\mcitedefaultmidpunct}
{\mcitedefaultendpunct}{\mcitedefaultseppunct}\relax
\EndOfBibitem
\bibitem[J\'o{}nsson(2010)]{jon10}
H.~J\'o{}nsson, \emph{Proc. Nat. Acad. Sci. U.S.A.}, 2010, \textbf{108},
  944--949\relax
\mciteBstWouldAddEndPuncttrue
\mciteSetBstMidEndSepPunct{\mcitedefaultmidpunct}
{\mcitedefaultendpunct}{\mcitedefaultseppunct}\relax
\EndOfBibitem
\bibitem[Meisner \emph{et~al.}(2011)Meisner, Rommel, and K\"astner]{mei11}
J.~Meisner, J.~B. Rommel and J.~K\"astner, \emph{J. Comput. Chem.}, 2011,
  \textbf{32}, 3456\relax
\mciteBstWouldAddEndPuncttrue
\mciteSetBstMidEndSepPunct{\mcitedefaultmidpunct}
{\mcitedefaultendpunct}{\mcitedefaultseppunct}\relax
\EndOfBibitem
\bibitem[Goumans and K\"astner(2011)]{gou11a}
T.~P.~M. Goumans and J.~K\"astner, \emph{J. Phys. Chem. A}, 2011, \textbf{115},
  10767\relax
\mciteBstWouldAddEndPuncttrue
\mciteSetBstMidEndSepPunct{\mcitedefaultmidpunct}
{\mcitedefaultendpunct}{\mcitedefaultseppunct}\relax
\EndOfBibitem
\bibitem[Einarsd\'o{}ttir \emph{et~al.}(2012)Einarsd\'o{}ttir, Arnaldsson,
  \'O{}skarsson, and J\'o{}nsson]{ein11}
D.~M. Einarsd\'o{}ttir, A.~Arnaldsson, F.~\'O{}skarsson and H.~J\'o{}nsson,
  \emph{Lect. Notes Comput. Sci.}, 2012, \textbf{7134}, 45\relax
\mciteBstWouldAddEndPuncttrue
\mciteSetBstMidEndSepPunct{\mcitedefaultmidpunct}
{\mcitedefaultendpunct}{\mcitedefaultseppunct}\relax
\EndOfBibitem
\bibitem[Rommel \emph{et~al.}(2012)Rommel, Liu, Werner, and K\"astner]{rom12}
J.~B. Rommel, Y.~Liu, H.-J. Werner and J.~K\"astner, \emph{J. Phys. Chem. B},
  2012, \textbf{116}, 13682\relax
\mciteBstWouldAddEndPuncttrue
\mciteSetBstMidEndSepPunct{\mcitedefaultmidpunct}
{\mcitedefaultendpunct}{\mcitedefaultseppunct}\relax
\EndOfBibitem
\bibitem[Kryvohuz and Marcus(2012)]{kry12}
M.~Kryvohuz and R.~A. Marcus, \emph{J. Chem. Phys}, 2012, \textbf{137},
  134107\relax
\mciteBstWouldAddEndPuncttrue
\mciteSetBstMidEndSepPunct{\mcitedefaultmidpunct}
{\mcitedefaultendpunct}{\mcitedefaultseppunct}\relax
\EndOfBibitem
\bibitem[K\"astner(2013)]{kae13}
J.~K\"astner, \emph{Chem. Eur. J.}, 2013, \textbf{19}, 8207--8212\relax
\mciteBstWouldAddEndPuncttrue
\mciteSetBstMidEndSepPunct{\mcitedefaultmidpunct}
{\mcitedefaultendpunct}{\mcitedefaultseppunct}\relax
\EndOfBibitem
\bibitem[\'Alvarez-Barcia \emph{et~al.}(2014)\'Alvarez-Barcia, Flores, and
  K\"astner]{alv14}
S.~\'Alvarez-Barcia, J.~R. Flores and J.~K\"astner, \emph{J. Phys. Chem. A},
  2014, \textbf{118}, 78\relax
\mciteBstWouldAddEndPuncttrue
\mciteSetBstMidEndSepPunct{\mcitedefaultmidpunct}
{\mcitedefaultendpunct}{\mcitedefaultseppunct}\relax
\EndOfBibitem
\bibitem[K\"astner(2014)]{kae14}
J.~K\"astner, \emph{WIREs Comput. Mol. Sci.}, 2014, \textbf{4}, 158\relax
\mciteBstWouldAddEndPuncttrue
\mciteSetBstMidEndSepPunct{\mcitedefaultmidpunct}
{\mcitedefaultendpunct}{\mcitedefaultseppunct}\relax
\EndOfBibitem
\bibitem[Kryvohuz(2014)]{kry14}
M.~Kryvohuz, \emph{J. Phys. Chem. A}, 2014, \textbf{118}, 535--544\relax
\mciteBstWouldAddEndPuncttrue
\mciteSetBstMidEndSepPunct{\mcitedefaultmidpunct}
{\mcitedefaultendpunct}{\mcitedefaultseppunct}\relax
\EndOfBibitem
\bibitem[Zheng and Truhlar(2010)]{zhe10}
J.~Zheng and D.~G. Truhlar, \emph{Phys. Chem. Chem. Phys.}, 2010, \textbf{12},
  7782--7793\relax
\mciteBstWouldAddEndPuncttrue
\mciteSetBstMidEndSepPunct{\mcitedefaultmidpunct}
{\mcitedefaultendpunct}{\mcitedefaultseppunct}\relax
\EndOfBibitem
\bibitem[Adler \emph{et~al.}(2007)Adler, Knizia, and Werner]{adler:07}
T.~B. Adler, G.~Knizia and H.-J. Werner, \emph{J. Chem. Phys.}, 2007,
  \textbf{127}, 221106\relax
\mciteBstWouldAddEndPuncttrue
\mciteSetBstMidEndSepPunct{\mcitedefaultmidpunct}
{\mcitedefaultendpunct}{\mcitedefaultseppunct}\relax
\EndOfBibitem
\bibitem[Knizia \emph{et~al.}(2009)Knizia, Adler, and Werner]{knizia:09}
G.~Knizia, T.~B. Adler and H.-J. Werner, \emph{J. Chem. Phys.}, 2009,
  \textbf{130}, 054104\relax
\mciteBstWouldAddEndPuncttrue
\mciteSetBstMidEndSepPunct{\mcitedefaultmidpunct}
{\mcitedefaultendpunct}{\mcitedefaultseppunct}\relax
\EndOfBibitem
\bibitem[Peterson \emph{et~al.}(2008)Peterson, Adler, and Werner]{peterson:08}
K.~A. Peterson, T.~B. Adler and H.-J. Werner, \emph{J. Chem. Phys.}, 2008,
  \textbf{128}, 084102\relax
\mciteBstWouldAddEndPuncttrue
\mciteSetBstMidEndSepPunct{\mcitedefaultmidpunct}
{\mcitedefaultendpunct}{\mcitedefaultseppunct}\relax
\EndOfBibitem
\bibitem[Werner \emph{et~al.}(2012)Werner, Knowles, Knizia, Manby,
  {Sch\"{u}tz}, Celani, Gy\"orffy, Kats, Korona, Lindh, Mitrushenkov, Rauhut,
  Shamasundar, Adler, Amos, Bernhardsson, Berning, Cooper, Deegan, Dobbyn,
  Eckert, Goll, Hampel, Hesselmann, Hetzer, Hrenar, Jansen, K\"oppl, Liu,
  Lloyd, Mata, May, McNicholas, Meyer, Mura, Nicklass, O'Neill, Palmieri, Peng,
  Pfl\"uger, Pitzer, Reiher, Shiozaki, Stoll, Stone, Tarroni, Thorsteinsson,
  and Wang]{MOLPRO}
H.-J. Werner, P.~J. Knowles, G.~Knizia, F.~R. Manby, M.~{Sch\"{u}tz},
  P.~Celani, W.~Gy\"orffy, D.~Kats, T.~Korona, R.~Lindh, A.~Mitrushenkov,
  G.~Rauhut, K.~R. Shamasundar, T.~B. Adler, R.~D. Amos, A.~Bernhardsson,
  A.~Berning, D.~L. Cooper, M.~J.~O. Deegan, A.~J. Dobbyn, F.~Eckert, E.~Goll,
  C.~Hampel, A.~Hesselmann, G.~Hetzer, T.~Hrenar, G.~Jansen, C.~K\"oppl,
  Y.~Liu, A.~W. Lloyd, R.~A. Mata, A.~J. May, S.~J. McNicholas, W.~Meyer, M.~E.
  Mura, A.~Nicklass, D.~P. O'Neill, P.~Palmieri, D.~Peng, K.~Pfl\"uger,
  R.~Pitzer, M.~Reiher, T.~Shiozaki, H.~Stoll, A.~J. Stone, R.~Tarroni,
  T.~Thorsteinsson and M.~Wang, \emph{MOLPRO, version 2012, a package of ab
  initio programs}, 2012, see https://www.molpro.net/\relax
\mciteBstWouldAddEndPuncttrue
\mciteSetBstMidEndSepPunct{\mcitedefaultmidpunct}
{\mcitedefaultendpunct}{\mcitedefaultseppunct}\relax
\EndOfBibitem
\bibitem[Zhao and Truhlar(2006)]{zhao:06}
Y.~Zhao and D.~G. Truhlar, \emph{J. Chem. Phys.}, 2006, \textbf{125},
  194101\relax
\mciteBstWouldAddEndPuncttrue
\mciteSetBstMidEndSepPunct{\mcitedefaultmidpunct}
{\mcitedefaultendpunct}{\mcitedefaultseppunct}\relax
\EndOfBibitem
\bibitem[Valiev \emph{et~al.}(2010)Valiev, Bylaska, Govind, Kowalski,
  Straatsma, van Dam, Wang, Nieplocha, Apra, Windus, and de~Jong]{nwchem}
M.~Valiev, E.~J. Bylaska, N.~Govind, K.~Kowalski, T.~P. Straatsma, H.~J.~J. van
  Dam, D.~Wang, J.~Nieplocha, E.~Apra, T.~L. Windus and W.~A. de~Jong,
  \emph{Comput. Phys. Commun.}, 2010, \textbf{181}, 1477--1489\relax
\mciteBstWouldAddEndPuncttrue
\mciteSetBstMidEndSepPunct{\mcitedefaultmidpunct}
{\mcitedefaultendpunct}{\mcitedefaultseppunct}\relax
\EndOfBibitem
\end{mcitethebibliography}
\providecommand*{\mcitethebibliography}{\thebibliography}
\csname @ifundefined\endcsname{endmcitethebibliography}
{\let\endmcitethebibliography\endthebibliography}{}

\end{document}